# Type-II Dirac Nodal Lines in double-kagome-layered CsV8Sb12

Yongqing Cai[1*], Jianfeng Wang[2,3*#], Yuan Wang[1*], Zhanyang Hao[1*], Yixuan Liu[1*], Zhicheng Jiang[4], Xuelei Sui[3], Xiao-Ming Ma[1], Chengcheng Zhang[1], Zecheng Shen[1], Yichen Yang[4], Wanling Liu[4], Qi Jiang[4], Zhengtai Liu[4], Mao Ye[4], Dawei Shen[4], Yi Liu[5], Shengtao Cui[5], Le Wang[1], Cai Liu[1], Junhao Lin[1], Bing Huang[3], Jia-Wei Mei[1#] and Chaoyu Chen[1#]

[1] Shenzhen Institute for Quantum Science and Engineering (SIQSE) and Department of Physics, Southern University of Science and Technology (SUSTech), Shenzhen 518055, China.

[2] School of Physics, Beihang University, Beijing 100191, China.

[3] Beijing Computational Science Research Center, Beijing 100193, China.

[4] State Key Laboratory of Functional Materials for Informatics and Center for Excellence in Superconducting Electronics, Shanghai Institute of Microsystem and Information Technology, Chinese Academy of Sciences, Shanghai 200050, China.

[5] National Synchrotron Radiation Laboratory, University of Science and Technology of China, Hefei, Anhui 230029, China.

[*] These authors contributed equally to this work.

[#] Correspondence should be addressed to J.W. (wangjf06@buaa.edu.cn), J.M. (meijw@sustech.edu.cn) and C.C. (chency@sustech.edu.cn)

Lorentz-violating type-II Dirac nodal line semimetals (DNLSs) [1-4], hosting curves of band degeneracy formed by two dispersion branches with the same sign of slope, represent a novel state of matter. While being studied extensively in theory [1-6], convincing experimental evidences of type-II DNLSs remain elusive. Recently, Vanadium-based kagome materials have emerged as a fertile ground to study the interplay between lattice symmetry and band topology [7-14]. In this work, we study the low-energy band structure of double-kagome-layered $CsV_8Sb_{12}$ and identify it as a scarce type-II DNLS protected by mirror symmetry. We have observed multiple DNLs consisting of type-II Dirac cones close to or almost at the Fermi level via angle-resolved photoemission spectroscopy (ARPES). First-principles analyses show that spin-orbit coupling only opens a small gap, resulting in effectively gapless ARPES spectra, yet generating large spin Berry curvature. These type-II DNLs, together with the interaction between a low-energy van Hove singularity and quasi-1D band as we observed in the same material, suggest $CsV_8Sb_{12}$ as an ideal platform for exploring novel transport properties such as chiral anomaly [15], Klein tunneling [16] and fractional quantum Hall effect [17-19].

In three-dimensional materials, the band crossing between valence and conduction bands can form discrete points, described as Dirac semimetals and Weyl semimetals[20]. The bands can also cross along a curve, described as nodal-line semimetals (NLSs)[21,22]. In general, NLSs can be categorized into type-I and type-II[3,4], according to the band dispersion slope around the band crossing. Compared with type-I NLSs, type-II NLSs host distinct properties, *e.g.*, chiral anomaly[15], and Klein tunneling[16]. Although quite a mount of theoretical efforts have been made on the prediction of type-II NLSs[1-6], the direct spectral evidence to realize the type-II NLS phase in these materials is still missing.

$A$V$_3$Sb$_5$ ($A$=K, Rb, Cs) family of materials intertwine charge density wave (CDW), superconductivity and band topology, exhibiting novel properties such as pair density wave[12], giant anomalous Hall conductivity[11], time reversal symmetry breaking (TRSB) and possible chiral flux phase[13]. Topological band features, such as van Hove singularity (VHS) and Dirac nodal loop[23,24], are observed and fundamental to shape those properties. However, despite intensive research efforts, there are still controversies concerning the existence of chiral flux phase, the chirality of CDW and its relation with TRSB[25,26]. Recent efforts[27,28] have succeeded in synthesizing doulbe-kagome-layered materials CsV$_8$Sb$_{12}$. Interestingly, neither CDW nor superconductivity is present in the normal states of single crystals, providing valuable chances of designing control experiment to examine the above controversies. To date, the band structure topology of these double-kagome-layered compounds remains to be clarified experimentally, especially for CsV$_8$Sb$_{12}$.

In this work, we identify double-kagome-layered CsV$_8$Sb$_{12}$ as a type-II DNLS. Combining photon-energy dependent ARPES measurement and density functional theory (DFT) analyses, we have observed multiple groups of type-II DNLs extending along $k_z$ direction, which are protected by mirror symmetry and lying close to or almost at the Fermi level. According to DFT, spin-orbit coupling (SOC) only opens a very small gap at the Dirac points, effectively yielding gapless Dirac point in the ARPES spectra yet large spin Berry curvature. We have also observed low-energy VHS interacting with a quasi-one-dimensional (quasi-1D) band, both of them featuring enhanced density of state. Our identification of type-II DNLs provides a clear picture to categorize the band structure topology of double-kagome-layered CsV$_8$Sb$_{12}$. The observation of these intrinsic, low-energy electronic features also stimulates further transport exploration for novel topological and correlated physic[15-19] in V-based kagome materials.

CsV$_8$Sb$_{12}$ adopts an orthorhombic lattice with space group of $Fmmm$[27,28]. Both $A$V$_3$Sb$_5$ and CsV$_8$Sb$_{12}$ share a common V$_3$Sb$_5$ unit that consists of a V-based kagome layer V$_3$Sb sandwiched by two Sb$_2$ honeycomb lattices. While $A$V$_3$Sb$_5$ has one V$_3$Sb$_5$ unit, CsV$_8$Sb$_{12}$ has two V$_3$Sb$_5$ units, plus an orthorhombic V$_2$Sb$_2$ layer in between (Fig. 1(a))[27,28]. Considering the mirror symmetry in the $z$ direction, the effective unit cell (red dashed box in Fig. 1(a)) with $c \sim 18.1$ Å consists of two {half Cs-V$_3$Sb$_5$-half V$_2$Sb$_2$} units (blue dashed box) that are related to each other by glide plane. This sublattice symmetry leads to an apparent $k_z$ periodicity of $4\pi/c$ as shown in Fig. 3(a).

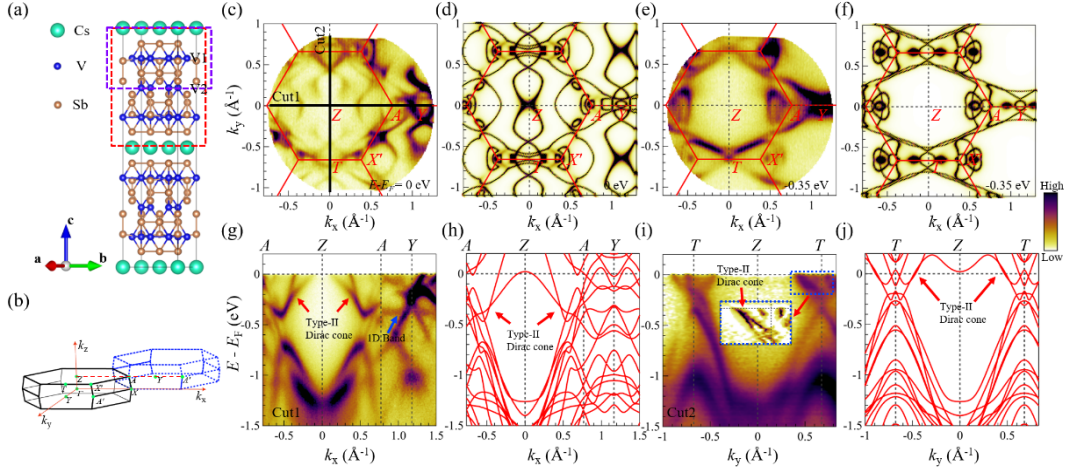

**Fig. 1. General electronic structure of CsV₈Sb₁₂.** (a) Lattice structure of CsV$_8$Sb$_{12}$ conventional cell. Cs, V and Sb atoms are represented by green, blue and brown balls respectively, where the V atoms in the kagome V$_3$Sb layer and orthorhombic V$_2$Sb$_2$ layer are labeled as V1 and V2. (b) First (solid black lines) and second (dashed blue lines) Brillioun zones (BZs) of CsV$_8$Sb$_{12}$ primitive cell. (c-f) Constant energy contours (CECs) at $E_F$ and 0.35 $eV$ below $E_F$ from both ARPES and DFT. (g-j) ARPES spectra and DFT calculated dispersions (without SOC) along high-symmetry paths, which are marked by solid black lines (Cut1 and Cut2) in (c). ARPES data here is measured with 61 eV photons ($k_z = 11.5 \times 2\pi/c$).

With the first and second Brillouin zones (BZs) shown in Fig. 1(b), Figs. 1(c-j) demonstrate the measured and calculated Fermi surfaces and band dispersions of CsV$_8$Sb$_{12}$. Satisfactory agreements between ARPES measurements and DFT calculations are reached concerning the general symmetry and detailed band structure. Because of the orthorhombic V$_2$Sb$_2$ layer, the symmetry of CsV$_8$Sb$_{12}$ is reduced to $D_{2h}$, in contrast to the $D_{6h}$ symmetry in $A$V$_3$Sb$_5$. In line to this, the Fermi surface and CEC (Figs. 1(c-f)) both show two-fold symmetry, with spectral features symmetric to $Z - A$ and $Z - T$ lines. ARPES spectra and DFT dispersions along these two high-symmetry paths are shown in Figs. 1(g-j) for comparison. The ARPES-measured electronic features of CsV$_8$Sb$_{12}$ can be roughly categorized into three groups: multiple (gapped) Dirac cones and VHS around $A$ and $T(Y)$ points, nearly flat bands along $A - Y$ (quasi-1D band) at ~0.3 $eV$ below $E_F$ (blue arrows) and type-II Dirac cones along $Z - A$ and $Z - T$ paths (red arrows). The multiple (gapped) Dirac cones and VHS are common featues found in kagome lattices, as already discussed in $A$V$_3$Sb$_5$[7-10]. Outstandingly, the multiple type-II Dirac cones are unique electronic features found in double-kagome-layered CsV$_8$Sb$_{12}$, as they are absent in single-kagome-layered $A$V$_3$Sb$_5$. Next, we first focus on the analyses of type-II Dirac cones, which will evolve into type-II DNLs along $k_z$, leaving the discussions of VHS and quasi-1D band later.

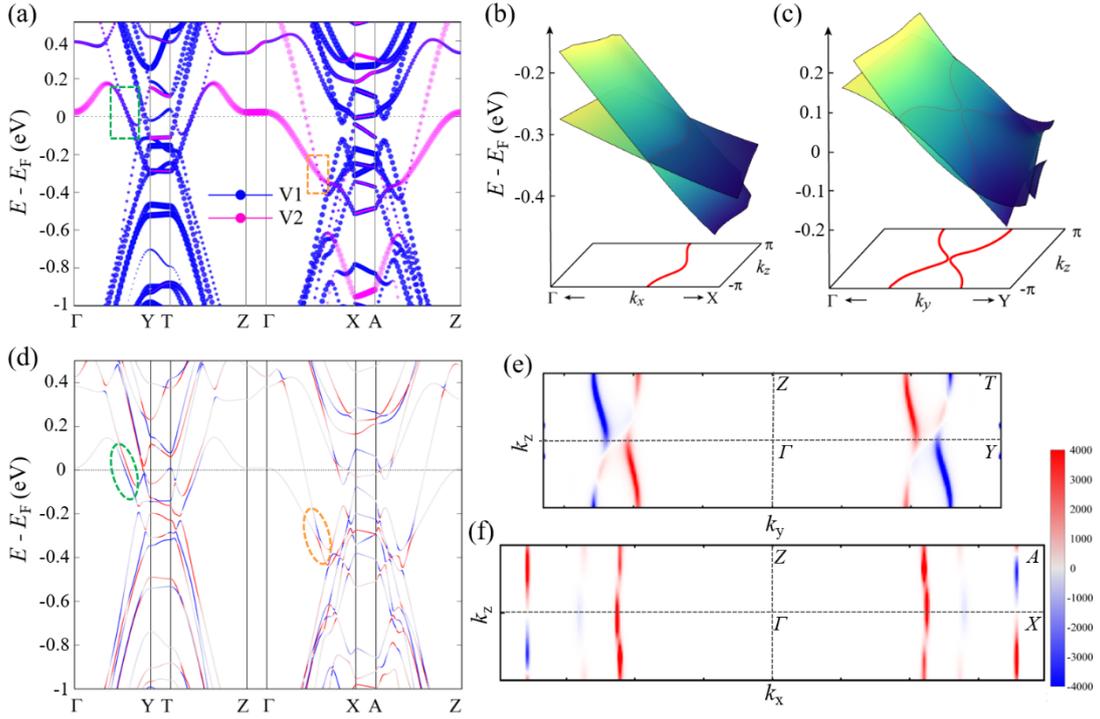

**Fig. 2. Type-II DNLs and contributions to spin Berry curvature by DFT calculations.** (a) Calculated band structure (without SOC) with $d$ orbital projections of kagome V1 (blue) and orthorhombic V2 (magneta) atoms. (b, c) 3D plot of 2D bands for the type-II DNLs highlighted in (a) by the orange and green dashed boxes for (b) and (c) respectively. Band crossings are projected to the 2D $k_x$-$k_z$ and $k_y$-$k_z$ planes (red lines). (d) Band structure with SOC effect. The color denotes the spin Berry curvature $\Omega_n^z(\boldsymbol{k})$ of the $n$th band for each $k$ point which ranges from negative (blue) to positive (red). Orange and green circles highlight the gap opening and spin Berry curvatures around the type-II DNLs. (e, f) $\boldsymbol{k}$-resolved spin Berry curvature after summation of the occupied bands in the $k_x = 0$ and $k_y = 0$ planes (2D BZ). The color bar is in arbitrary units.

As seen in Figs. 1(g-j), type-II Dirac cones, *i.e.*, tilted Dirac cone containing two dispersion branches with the same sign of slope, are found along both $Z - A$ and $Z - T$ lines. The Dirac energies are found $E_D \sim -0.2 \ eV$ for the one along $Z - A$ and $E_D \sim 0 \ eV$ for that along $Z - T$. Fig. 2(a) shows that the bands near $E_F$ are mainly contributed by the $d$ orbitals of kagome V1 atoms and orthorhombic V2 atoms. Especially, the type-II Dirac cones along $Z - A$ are dominated by the $d$ band crossings of V2 atoms. From Fig. 2(a), the type-II Dirac cones also exist along $\Gamma - X$ and $\Gamma - Y$, indicating the possibility of type-II DNLs along $k_z$. To demonstrate this, we plot the 2D band structures near the band crossings with $k_z$ changed from $-\pi/c$ to $\pi/c$ in Figs. 2(b) and (c). Two groups of NLs are formed in the $k_x$-$k_z$ and $k_y$-$k_z$ planes, which include one and two lines along $k_z$, respectively. With opposite eigenstate parities of $M_y$ ($M_x$) for these two crossed bands in $k_x$-$k_z$ ($k_y$-$k_z$) plane, these type-II DNLs are protected by mirror symmetry.

It is noted that the above results are based on the calculations without SOC effect in terms of single group representation. Taking SOC into account, the DNLs will open an energy gap (< 30 meV), as shown in Fig. 2(d). Considering the feature of type-II band crossing, this gap is still smaller than the energy distribution curve (EDC) width of ARPES spectra, resulting in effectively gapless features as observed. As we know, the SOC-induced Dirac gap will produce a large Berry curvature or spin Berry curvature. This is confirmed by our spin Berry curvature calculations (see Method section) as shown in Fig. 2(d). The band anti-crossing along NLs by SOC will give a particularly large contribution. As shown in Figs. 2(e) and (f), the hot spot of spin Berry curvature distribution in the 2D BZ ($k_x = 0$ and $k_y = 0$ planes) well follows the shape of NLs. Besides the type-II NLs, other band anti-crossings approaching to $X - A$ contribute to another smaller hot spot (Figs. 2(d) and (f)). The large spin Berry curvature by NLs should yield novel spin-electronic transport behaviours, $e.g.$, a large intrinsic spin Hall effect.

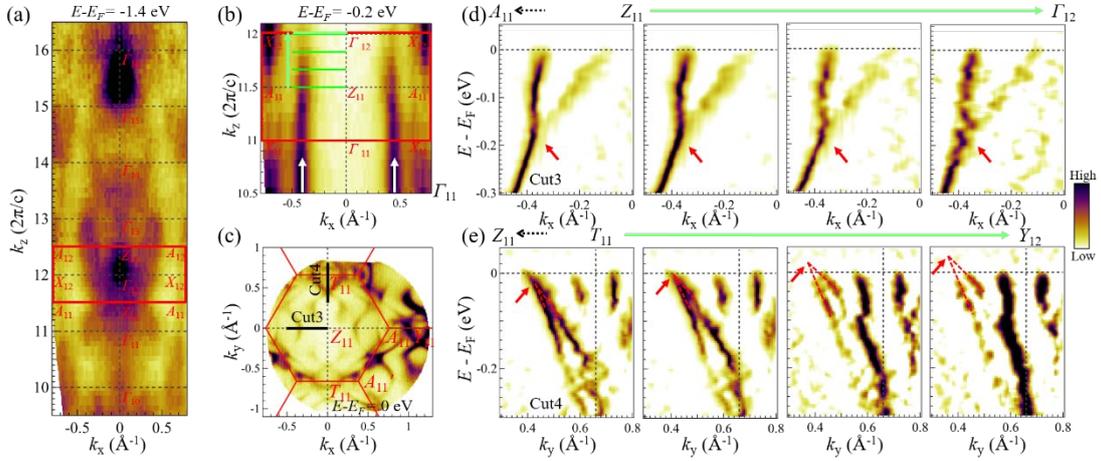

**Fig. 3. Type-II DNLs from ARPES measurements.** (a, b) Photon-energy dependent ARPES measurements of CECs in $k_z - k_x$ plane at $1.4$ $eV$ (a) and $0.2$ $eV$ (b) below $E_F$. (c) CEC in $k_x - k_y$ plane at $k_z = 11.5 \times 2\pi/c$. (d) ARPES spectra (2$^{rd}$ derivative) along $k_x$ to view the type-II Dirac cone at a series of $k_z$ values from $Z - A$ to $\Gamma - X$ (green lines in (b) and Cut3 in (c)). (e) Same as (d) but along $k_y$ at a series of $k_z$ values from $Z - T$ to $\Gamma - Y$ (Cut4 in (c)). Red arrows in (d) and (e) point out the type-II Dirac points. White arrows in (b) mark the NLs formed by type-II Dirac cones along $k_z$.

To experimentally prove the existence of these type-II DNLs, we have perfomed photon-energy dependent measurement. As exemplified in Figs. 3(a-b) for spectral intensity along $Z - A$, clear $k_z$ periodicity can be found from the CEC in $k_z - k_x$ space at $1.4$ $eV$ below $E_F$. This enables us to locate the bulk $\Gamma$ points in 3D BZ. Four $Z - A$ ARPES spectra from a series of $k_z$ values (green lines in Fig. 3(b) and Cut3 in Fig. 3(c)), covering a momentum space between bulk $Z_{11}$ to $\Gamma_{12}$, are shown in Fig. 3(d). Red arrows in Fig. 3(d) not only mark the gapless behaviour of these Dirac cones, but also their constant Dirac energies ($E_D \sim -0.2$ $eV$) along $k_z$, suggesting the formatioin of type-II DNLs. To visualize the NLs, the ARPES spectral intensity is integrated in a $0.05$ $eV$ enegy window around $E_D \sim -0.2$ $eV$ and plotted in $k_z - k_x$ space in Fig. 3(b). As marked by the white arrows, the straight features along $k_z$ directly visualize these NLs formed by type-II Dirac cones along $Z - A$. Similar analyses shown in Fig. 3(e) also demonstrate the existence of NLs around $E_F$ formed by type-II Dirac cones along $Z - T$. Along this direction, two bands contributed by the $d$ orbitals of the V1 atoms cross twice and form two Dirac nodes (Fig. 2(a)). At $k_z = 0$ plane, these two nodes almost merge together, while at $k_z = \pi$ plane, they are separated obviously (Fig. 2(c)). The Dirac cone as observed in Fig. 3(e) is supposed to be the lower half. The Dirac point moves upwards from $k_z = \pi$ plane to $k_z = 0$ plane, which is in good agreement with our DFT calculations (Fig. 2(c)). In particular, the Dirac point crosses $E_F$ along $k_z$, which should play an important role in shaping the transport behavior of the material.

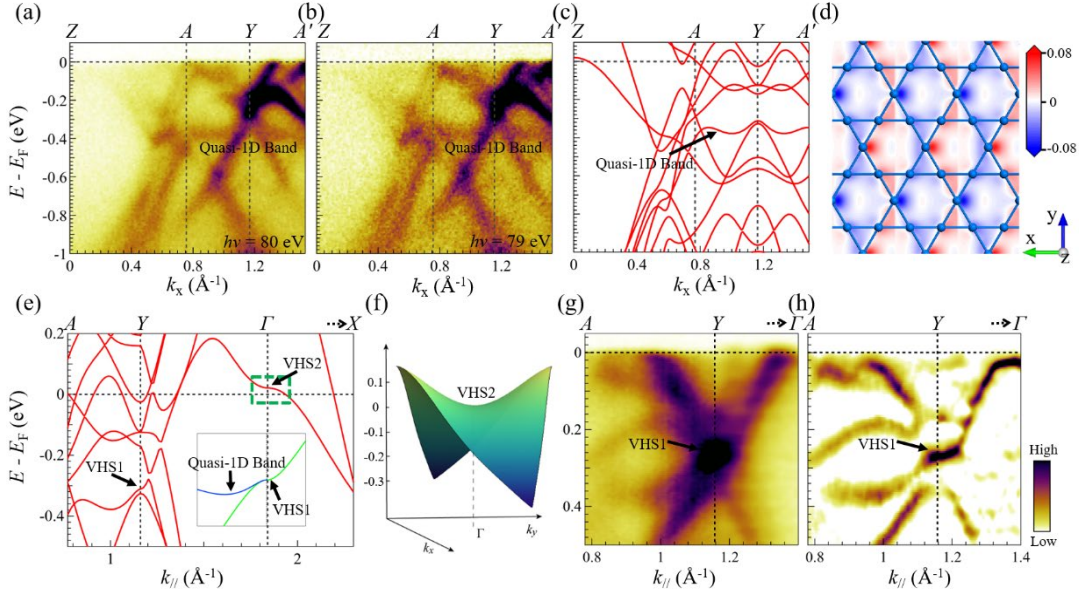

**Fig. 4. Interacted quasi-1D band and VHS in CsV₈Sb₁₂.** (a, b) ARPES spectra along $Z - A - Y - A'$ taken with photon energies of $80\ eV$ (a) and $79\ eV$ (b). (c) DFT calculated band structure along the same momentum paths. (d) Wave function plot of electronic states for quasi-1D band. (e) DFT calculated band along $A - Y - \Gamma - X$. Interacted VHS1 and quasi-1D band are highlighted in the inset and newly formed VHS2 is highlighted by a green dashed box. (f) 3D plot of VHS2 in the $k_x$-$k_y$ plane around $\Gamma$. (g, h) ARPES spectrum (g) and its second derivative image (h) along $A - Y - \Gamma$.

As one of the characteristic electronic features of kagome lattice, flat bands in single-kagome-layered CsV₃Sb₅ are located beyond $1\ eV$ below and above $E_F$ according to the DFT calculations[9], rendering their physical responses hardly observable. Interestingly, a low-energy flat band is observed in double-kagome-layered CsV₈Sb₁₂. As shown in Figs. 4(a) and (b), our ARPES spectra with both photon energies of $80\ eV$ and $79\ eV$ demonstrate a nearly non-dispersive band along the BZ boundary $A - Y - A'$ at $\sim 0.4\ eV$ below $E_F$. The excitation energy independence of this band reveals its intrinsic nature. The DFT calculated band structure along the same high-symmetry paths is shown in Fig. 4(c), which is in good agreement with the ARPES measurements. Obviously, a weakly dispersive band is present along $A - Y - A'$ but absent along other paths, $e.g.$, along $\Gamma - Y$ or $Z - T$ as shown in Fig. 4(e) or 2(a), indicating its quasi-1D behaviors. Our DFT calculations further reveal that this quasi-1D band is contributed by the $d$ orbitals of kagome V1 atoms. However, it is different from the usual flat band in the kagome lattice which arises from a destructive phase interference of hopping. To visualize the origin of this quasi-1D band, we plot the wave function of its electronic states in Fig. 4(d). Instead of uniformly distributing on three V atoms in the kagome lattice, the electronic states of quasi-1D band are mainly located at one V atom. The wave functions hybridize with the adjacent ones to form an approximate striped distribution. In addition, the $p$ orbitals of Sb atoms from orthorhombic $V_2Sb_2$ layer also have small contributions to the hybridizations of these states. Hence, this quasi-1D band in $CsV_8Sb_{12}$ originates from the combined interaction of the kagome $V_3Sb$ layer and orthorhombic $V_2Sb_2$ layer.

VHSs, another typical feature of kagome lattice, are also found in this double-kagome-layered system. As shown in Fig. 4(e), near $E_F$, DFT predicts two VHSs at $Y$ (VHS1) and $\Gamma$ (VHS2), respectively. Remarkably, the VHS2 at $\Gamma$ lies slightly above $E_F$ (Fig. 4(f)), temptingly calls for electron doping to manifest its physical response. It is noted that this VHS originates from the orthorhombic $V_2Sb_2$ layer rather than the kagome lattice (Fig. 2(a)). Considering that the VHS usually exists at BZ boundary, the VHS2 at BZ center is scarce and deserves further investigation. VHS1 at $Y$, $\sim 0.3 \ eV$ below $E_F$, is directly observed in ARPES spectra (Figs. 4(g-h)). More interestingly, it almost has the same binding energy with the quasi-1D band and they interact with each other at $Y$, as illustrated in the inset of Fig. 4(e). Since both quasi-1D and VHS feature enhanced density of states, as directly visualized by the very high intensity in Figs. 4(g-h), their low-energy nature calls for topological and correlated electronic transport experiment via local electrostatic gates.

Combining experimental and calculational analyses, we have performed systematic investigations on the electronic structure of double-kagome-layered $CsV_8Sb_{12}$. We have revealed unique electronic features in this compound, including multiple type-II DNLs as well as interacted quasi-1D band and VHS. Our results provide the first spectral evidence of type-II NLSM phase and offer a unique platform to investigate the distinct properities of these novel electronic features. The type-II DNLs, especially the one lying at the Fermi level, should play a significant role in shaping the transport response of the material and deserve further theoretical and experimental study. The zone-center VHS and enhanced spectral weight featured by interacted quasi-1D band and VHS stimulates local electrostatic control for topological and correlated physics.

## ACKNOWLEDGEMENTS

This work is supported by National Natural Science Foundation of China (NSFC) (Grants Nos. 12074163 and 12004030), the Guangdong Innovative and Entrepreneurial Research Team Program (Grants Nos. 2017ZT07C062 and 2019ZT08C044), Shenzhen Science and Technology Program (Grant No. KQTD20190929173815000), the University Innovative Team in Guangdong Province (No. 2020KCXTD001), Shenzhen Key Laboratory of Advanced Quantum Functional Materials and Devices (No. ZDSYS20190902092905285), Guangdong Basic and Applied Basic Research Foundation (No. 2020B1515120100), and China Postdoctoral Science Foundation (2020M682780). The authors acknowledge the assistance of SUSTech Core Research Facilities. The calculations were performed at Tianhe2-JK at Beijng Computational Science Research Center.

## Methods

**Sample growth and characterization.** High quality $CsV_8Sb_{12}$ single crystals were grown by self-flux method. High purity Cs(clump), V(powder) and Sb(ball) were mixed with a ratio of $1:6:18$ in the glovebox and placed into an alumina crucible. The crucible was then double sealed into the evacuated quartz tubes to prevent the quartz tube from being corroded by Cs during heating. The assembly was first heated very slowly to 500 °C and kept for 5 h. Then the furnace was raised to 1100 °C and held at this temperature for another 24 h for proper homogenization. Finally, the furnace was cooled down to 900 °C at a rate of 1 °C/h. The crystals were separated from the flux by centrifuging.

**ARPES measurement.** ARPES measurements were performed at the BL03U beamline of the Shanghai Synchrotron Radiation Facility (SSRF) and beamline 13U of the National Synchrotron Radiation Laboratory (NSRL). The energy resolution was set at 15 meV for Fermi surface mapping and 7.5 meV for band structure measurements. The angular resolution was set at 0.1°. Samples were cleaved *in situ* under ultra-high vacuum conditions with pressure better than $5 \times 10^{-11}$ mbar and temperatures below 20 K.

**First-principles calculations.** The first-principles calculations are performed using the Vienna ab initio simulation package[29] within the projector augmented wave method[30] and the generalized gradient approximation of the Perdew-Burke-Ernzerhof[31] exchange-correlation functional. The plane-wave basis with an energy cutoff of 400 eV, the experimental lattice constants of $a=b=5.495$

Å and $c$=9.308 Å, and the $\Gamma$-centered 9×9×9 $k$-point meshes are adopted. The spin polarized calculations are tested, which gives a nonmagnetic ground state. The SOC effect is also considered in part of our calculations. A tight-binding (TB) Hamiltonian based on the maximally localized Wannier functions (MLWF)[32] is constructed to get the energy eigenvalues and eigenstates for further Fermi surface plot using the WannierTools package[33]. Using MLWF, the spin Berry curvature for the $n$th band at $\boldsymbol{k}$ is calculated by

$$\Omega^c_{n,ab}(\boldsymbol{k}) = -\sum_{n'\neq n} \frac{2\text{Im}[\langle \boldsymbol{k}n|\hat{j}^c_a|\boldsymbol{k}n'\rangle\langle \boldsymbol{k}n'|\hat{v}_b|\boldsymbol{k}n\rangle]}{(\varepsilon_{\boldsymbol{k}n}-\varepsilon_{\boldsymbol{k}n'})^2}.$$

Here, $\hat{j}^c_a = \{\hat{s}_c, \hat{v}_a\}/2$ is the spin current operator, $a$, $b$ and $c$ denote the three Cartesian directions $x$, $y$, and $z$, $\hat{s}_c = (\hbar/2)\beta\Sigma^c$, where $\beta$ is a 4×4 matrix and $\Sigma^c$ is the spin operator in the Dirac equation, and $|\boldsymbol{k}n\rangle$ represents the periodic part of the Bloch wave function with energy $\varepsilon_{\boldsymbol{k}n}$.